\documentclass[aps,prapplied,amsmath,amssymb,reprint]{revtex4-2}
\usepackage{CJK}
\usepackage{dcolumn}
\usepackage{bm}
\usepackage{siunitx}
\usepackage{float}

\usepackage{fancyhdr}
\usepackage{pgfplots}
\usepackage{graphicx}
\pgfplotsset{compat=1.16}
\graphicspath{{figs/}} 

\usepackage{hyperref}
\hypersetup{
	colorlinks=true,
	linkcolor=blue,
	anchorcolor=blue,
	citecolor=blue,
        urlcolor=blue,
}

\newcommand {\dif}{\mathop{}\!\mathrm{d}}
\newcommand {\weim}{\,\mu\text{m}}
\newcommand {\weis}{\,\mu\text{s}}

\begin{document}

\begin{CJK*}{UTF8}{gbsn}
\preprint{11\#202B}
\title{Numerical simulation of the radiation force from transient acoustic fields: \texorpdfstring{\\}{}Application to laser-guided acoustic tweezers}

\author{Shuhan Chen 
}
\author{Qing Wang 
}
\author{Qi Wang 
}
\author{Jia Zhou 
}
\email{jia.zhou@fudan.edu.cn}
\author{Antoine Riaud}

\email{antoine\_riaud@fudan.edu.cn}
\affiliation{State Key Laboratory of ASIC and System, School of Microelectronics, Fudan University, Shanghai 200433, China}
\date{\today}

\begin{abstract}
  Using pulsed acoustic waves could provide a superior selectivity for microscale acoustic tweezers. However, the theory for the radiation force of pulsed acoustic waves has only been recently derived and no numerical implementations are available. In this paper, we present a finite-element implementation of this model to simulate the transient acoustic radiation force on small spheres. We use the model to simulate laser-guided acoustic tweezers and optimize their performance. By enabling numerical simulations of the transient radiation force, this work may accelerate the rational design of pulse-based high-selectivity acoustic tweezers devices.
\end{abstract}


\maketitle
\end{CJK*}
\section{\label{sec:intro}Introduction}
  The acoustic radiation force (ARF), a steady force created by large-amplitude acoustic waves, is a convenient means to achieve micro-object manipulation such as micro-sample separation \cite{li2015standing,ahmed2017acoustic,dow2018acoustic} and enrichment \cite{antfolk2015acoustofluidic}, cell sorting \cite{augustsson2016iso,schmid2014sorting}, and single-cell manipulation \cite{baudoin2020spatially}.
  
  Using transient excitation such as pulses could enable much more precise manipulation than when using time-periodic acoustic fields \cite{li2015standing,ahmed2017acoustic,dow2018acoustic,antfolk2015acoustofluidic,augustsson2016iso,schmid2014sorting,baudoin2020spatially}. First, pulsed acoustic manipulation is less disturbed by Rayleigh acoustic streaming \cite{hoyos2013controlling,castro2016study} because radiation force is established much faster than streaming \cite{goering2021dynamic,goering2022measuring}. Second, using acoustic wave packets allows to localize the acoustic interference pattern and therefore to control the spatial extent of the acoustic trapping region \cite{collins2016acoustic}. Indeed, standing waves exert radiation forces considerably larger than traveling waves (in the small particle limit), which allows to neglect the acoustic field outside the interference region. Laser-guided acoustic tweezers (LGAT) \cite{wang2022laser} use this interference principle to create a hybridized radiation force landscape that couples a high-amplitude piezo-generated acoustic field (strong, Z-field) acoustic field and a light-patterned photogenerated acoustic field (weak, L-field). The hybridized field retains the spatial information of the L-field and the strength of the Z-field. 
  
  Despite these potential applications, theoretical and numerical studies of transient acoustic fields are still rare. A crying example of the limited current understanding of transient nonlinear acoustics is the suppression of acoustic streaming by acoustic pulses \cite{hoyos2013controlling,castro2016study}, where the only model available for transient streaming \cite{muller2015theoretical} has been unable to qualitatively explain experimental observations \cite{goering2021dynamic,goering2022measuring}. Likewise, there are no numerical schemes to study transient ARF directly.
  
  In this paper, we implement the recent generalization of the radiation force theory for small spheres (Gor'kov theory \cite{gorkov1961forces}) to transient acoustic fields \cite{wang2021acoustic}. Besides requiring that the objects are spherical and much smaller than the acoustic wavelength at the considered bandwidth, this theory overlooks micro-streaming and would not be suitable for studying very dense particles in highly viscous fluids (such as copper in glycerol \cite{baasch2019acoustic}). It also requires that scattering events do not overlap and therefore extreme care must be taken when investigating the response of bubbles or other high-quality acoustic resonators. Nonetheless, the underpinning assumptions suggest that the theory is valid for cells or plastic particles immersed in water. 
   
  The manuscript is arranged as follows. After summarizing the theory of dynamic acoustic radiation force, section~\ref{sec:method} details a numerical implementation of the dynamic ARF and details the model setup for the simulation of an LGAT. Section~\ref{sec:result} briefly presents the acoustic field in an LGAT device and then analyzes the resulting ARF depending on the location of the particle and on the phase difference between the laser and the piezoelectric. Then, the excitation parameters of the LGAT (pulse duration and laser beam width) are optimized to maximize the ARF.
\section{\label{sec:method}Theories and Methods}
\subsection{Transient ARF: Governing equations}
Our simulations implement the theory of ARF on small rigid spheres in transient acoustic fields of Wang \textit{et al.} \cite{wang2021acoustic}. Similar to most other theories of radiation pressure, the wave peak pressure amplitude $p_{\mathrm{max}}$ is assumed to be small such that ${p_{\mathrm{max}}/{\rho_{0}c_{0}^{2}}} = \epsilon \ll 1$ where $\epsilon$ is the acoustic Mach number, with $\rho_{0}$ the quiescent fluid density. According to the perturbation method, the fluid quantities can be resolved as
\begin{subequations}
	\begin{align}
		p&=p_{0}+p_{1}+p_{2},\\
		\mathbf{v}&=\mathbf{v}_{0}+\mathbf{v}_{1}+\mathbf{v}_{2}, \label{eq:1st_ap} \\		
		\rho&=\rho_{0}+\rho_{1}+\rho_{2},
	\end{align}	
\end{subequations}
where subscripts $0-2$ indicate the perturbation order of each term $q_{0} = q_{1}O(\epsilon) = q_{2}O\left( \epsilon^{2} \right)$ where $q$ can be any physical field among pressure, density and velocity. Accordingly, time-invariant 0-order terms are interpreted as constant (hydrostatic) contributions, 1\textsuperscript{st} order terms as acoustic contributions, and 2\textsuperscript{nd} order terms as nonlinear effects such as radiation pressure or acoustic streaming. Although the theory of radiation force requires knowledge of the total acoustic field, Gor'kov theory is conveniently expressed using the incident acoustic field only. Henceforth, all acoustic quantities will refer to the incident field in the remaining of the paper.

Compared to Gor'kov theory, the assumption of single-frequency acoustic fields is extended to wave packets of finite duration $\tau$: all the wave quantities $q_1$ are required to keep the initial state and final state consistent, which means satisfying the condition $ q_1\left( 0,\mathbf{r}\right) = q_1\left( \tau,\mathbf{r}\right) $, where $\mathbf{r}$ denotes the position vector and the origin of time can be chosen arbitrarily. Additionally, pulse duration must be short enough to neglect the sphere displacement compared to the shortest wavelength, which can be evaluated by $1/\tau \gg f_{\mathrm{max}} \epsilon^{2}$, where $f_{\mathrm{max}}$ refers to the highest frequency in the pulse.

In the case of an inviscid fluid (viscoacoustic and thermoacoustic boundary layers much thinner than the particle diameter \cite{settnes2012forces,karlsen2015forces}), the acoustic radiation force reads: 
\begin{equation} 
	\mathbf{F}_{\mathrm{rad}} = - V_{\mathrm{p}}\nabla U,
\end{equation}
with the particle volume $V_{\mathrm{p}}=4\pi R_{\mathrm{p}}^{3}/3$ and the generalized Gor'kov potential: 
\begin{equation}
	U = \left\langle U_{\mathrm{inst}} \right\rangle = \frac{1}{t_\mathrm{tot}}{\int{U_\mathrm{inst} \dif t}}, \label{eq:u_avg}
\end{equation}
where $U_{inst}$ is interpreted as a fictive (but convenient) instantaneous Gor'kov potential: 
\begin{equation}
	U_{\mathrm{inst}} = f_{1}\mathcal{V} - \frac{3}{2}f_{2}\mathcal{K}. \label{eq:u_inst}
\end{equation}
Here $\mathcal{K} ={\rho_{0}\mathbf{v}_{1} \cdot \mathbf{v}_{1}}/2$ and $\mathcal{V}={ p_{1}^{2}/(2\rho_{0}c_{0}^{2})}$ represent instantaneous kinetic energy and potential energy respectively. $
f_{1}=1-\rho_{0}c_{0}/(\rho_{\mathrm{p}}c_{\mathrm{p}})$ is the monopole scattering coefficient and $
f_{2}=2\left(\rho_{\mathrm{p}}+\rho_{0}\right)/(2\rho_{\mathrm{p}}+ \rho_{0})$ is the dipole scattering coefficient. 

Therefore, computation of the radiation force requires knowing the incident acoustic field, which can be computed using transient acoustic pressure models:
\begin{equation}
	\frac{1}{c_{0}^{2}}\frac{\partial^{2}p_{1}}{\partial t^{2}} + \nabla^{2}p_{1} = 0,
\end{equation}
from which the fluid acceleration can be deduced using the Euler equation:
\begin{equation}
    \mathbf{a}_{1} = - \frac{1}{\rho_{0}}\nabla p_{1},
\end{equation}
and the velocity:
\begin{equation}
	\mathbf{v}_{1} = {\int{\mathbf{a}_{1}\dif t}}. \label{eq:1st_v}
\end{equation}
Under these hypotheses, the general Gor'kov approach can be used for calculating ARF in such transient situations. 

\subsection{Transient ARF: Numerical implementation}
The numerical model for the simulation of hybridized acoustic fields was configured in the commercial finite-element-method software COMSOL Multiphysics\textsuperscript{\textregistered} version 5.4, and the parameters used are given in Table~\ref{tab:parameter}. The model uses the ``Pressure Acoustics, Transient'' module to solve the first-order acoustic pressure field in the fluid domain (Eq.~\eqref{eq:1st_ap} and to deduce the acceleration and velocity fields with Eq.~\eqref{eq:1st_v}). The settings for transient wave solving follow the approach suggested by Ref.~\cite{comsol1118resolving}.

Transient acoustic field studies are necessarily associated with bandwidth issues. On the one hand, the excitation signal (such as a short laser pulse) may have a very wide bandwidth. On the other hand, integration of the acoustic partial differential equation is often done with a constant time step and the grid imposes constraints on the minimum wavelength that can be resolved (that is, the maximum frequency). The Courant-Friedrichs-Lewy (CFL) condition provides guidelines to set the time step ${\Delta}t$ depending on the maximum mesh size $h$ \cite{muller2015theoretical,comsol1118resolving}: 
\begin{equation}
	\mathrm{CFL} = \frac{c\Delta t}{h},
\end{equation}
where $c$ is the wave speed. Using the default second-order (quadratic) mesh elements, the recommended CFL number is approximately 0.1. Thus, given a maximum frequency $f_{\text{max}}$ that needs to be resolved, the CFL yields the time step ${\Delta}t =\mathrm{CFL} / (Nf_{\text{max}})$ where $N$ is the number of elements per local wavelength $\lambda$. Due to the period doubling of the Gor'kov potential (terms in $p_{1}^2$ and $v_{1}^2$), we choose $N=12$ instead of the more common $N=6$ in linear acoustics.

Boundary conditions must be designed carefully according to this bandwidth constraint. The most straightforward is to directly set an acceleration that fits in the simulated frequency bandwidth as the boundary condition. However, in some cases, the velocity needs to be specified instead (for instance when the velocity is known from laser Doppler vibrometer measurements) which can yield to step-wise acceleration if there is a slope-break in the velocity. To address this problem, the input velocity is first multiplied by a smoothing window, and then derived to obtain the acceleration boundary condition.

The velocity field is computed (Eq.~\eqref{eq:1st_v}) by coupling the pressure acoustic model to a ``Distributed Ordinary Differential Equation (DODE)'' module in COMSOL with the damping factor set to 1 in DODE. 

The first-order velocity is then used to calculate the instantaneous Gor'kov potential $U_{\mathrm{inst}}$ via Eq.~\eqref{eq:u_inst} along with the first-order acoustic pressure. Similarly, Eq.~\eqref{eq:u_avg} is implemented in the ``Distributed Ordinary Differential Equation (DODE)'' module to obtain $U$. Finally, the ARF is obtained as the gradient of $U$ (using the COMSOL function \verb|diff()|). Given the small size of the geometry, computer memory is not a constraint in this study, and all the fields are computed using a single fully-coupled time-dependent solver for simplicity, but the perturbation approach of the model could allow using a segregated solver that solves the fields one after another for each time step.

\begin{table}[hbtp]
	\caption{\label{tab:parameter}
		Summary of simulation parameters. }
	\begin{ruledtabular}
		\begin{tabular}{lcc}
			\multicolumn{3}{l}{\textit{Electroacoustic wave parameters}} \\
			Frequency & $f^{\text{Z}}$ & \SI{10}{\MHz} \\
			Velocity magnitude & $v_{0}^{\text{Z}}$ & \SI{1.83}{\m/s} \\
			Number of shots & $n^{\text{Z}}$ & 5 \\
			Density	& $\rho_{\text{z}}$ & \SI{4640}{\kg\per\m^{3}} \\
			\multicolumn{3}{l}{\textit{Laser-induced acoustic wave parameters}} \\
			Frequency & $f^{\text{L}}$ & \SI{10}{\MHz} \\
			Velocity magnitude & $v_{0}^{\text{L}}$ & \SI{0.6}{\m/s} \\
			Radius of spot waist & $R_{\text{w}}$ & $20\weim$\\
			\multicolumn{3}{l}{\textit{Fluid}} \\		
			Density & $\rho_{0}$ & \SI{1050}{\kg\per\m^{3}} \\
			Speed of sound & $c_{0}$ & \SI{1500}{\m/s} \\
			Domain width (radius) & $R_{\text{d}}$ & $375\weim$ \\
			Domain height & $h_{\text{d}}$ & $200\weim$ \\
			\multicolumn{3}{l}{\textit{Particle}} \\
			Density & $\rho_{\text{p}}$ & \SI{1050}{\kg\per\m^{3}} \\ 
			Radius & $R_{\text{p}}$ & $10\weim$ \\
			\multicolumn{3}{l}{\textit{Mesh} \& \textit{Solver}} \\		
			CFL number & CFL & 0.5 \\
			Units per local wavelength & $N$ & 12 \\
			\multicolumn{3}{l}{\textit{Others}} \\		
			Simulation time & $t_{\text{sim}}$ & $2\weis$ \\
			Duration between pulse series & $t_{\text{tot}}$ & $20\weis$ \\
		\end{tabular}
	\end{ruledtabular}
\end{table}

\subsection{Simulation of an LGAT: Initial and boundary conditions}

\begin{figure}[htbp]
	\includegraphics[width=\columnwidth]{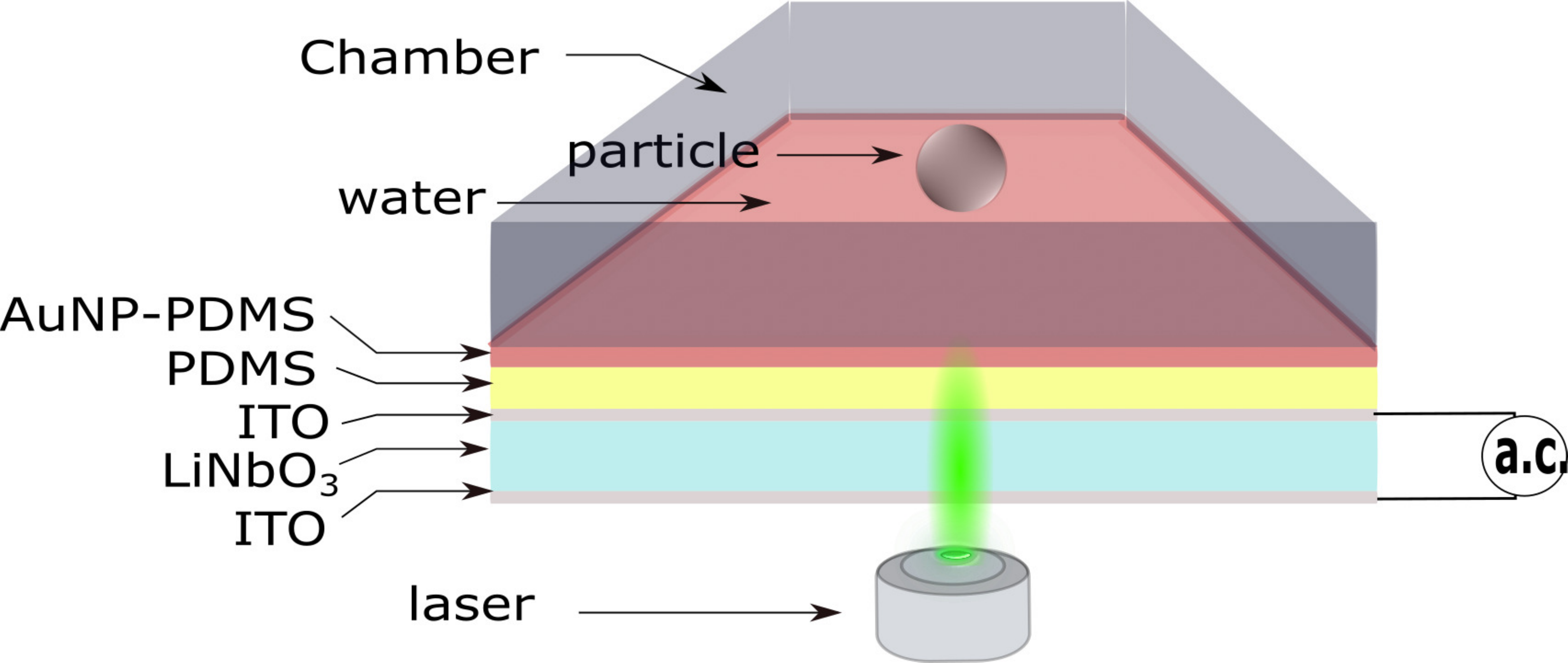}
	\caption{\label{fig:experiment} Laser-guided acoustic tweezers (LGAT) structure. The particle is manipulated by the combination of a weak and a strong acoustic field. The interference between the two fields yields a radiation force with spatial characteristics similar to the weak field but with a much higher amplitude. The L-field is generated by the photoacoustic conversion of a pulsed laser beam on the gold nanoparticle-polydimethylsiloxane (AuNP-PDMS) composite. The composite selectively absorbs green light and is transparent to other wavelengths. The Z-field is generated by piezoelectric conversion of an electric signal (alternating current, a.c.) in the LiNbO\textsubscript{3}. The electrodes (indium tin oxide, ITO) are transparent. As a result, the whole device is transparent to most optical wavelengths except green, and the particles under manipulation can be directly visualized.}
\end{figure}

In an LGAT (schematic shown in Fig.~\ref{fig:experiment}), two acoustic sources interfere to generate an acoustic trap. A piezoelectric acoustic source (strong field, labeled with the superscript ``Z'') yields a plane traveling wave of high amplitude which creates no in-plane acoustic radiation force due to symmetry constraints (the wave is in-plane invariant). This wave interferes with a weaker acoustic wave (weak field, labeled with the superscript ``L'') generated by the photoacoustic conversion of a laser pulse. The combined incident pressure and velocity fields read $p_{1} = p_{1}^{\mathrm{L}} + p_{1}^{\mathrm{Z}}$ and $\mathbf{v}=\mathbf{v}_{1}^\mathrm{L}+\mathbf{v}_{1}^\mathrm{Z}$, respectively. According to Eqs.~\eqref{eq:u_avg} and \eqref{eq:u_inst}, this yields a Gor'kov potential with cross terms $U=U^\mathrm{ZZ}+U^\mathrm{ZL}+U^\mathrm{LL}$. The vanishing gradient of spatially invariant $U^\mathrm{ZZ}$ leads to a negligible in-plane force, while $U^\mathrm{LL}$ is negligible due to the small photoacoustic conversion efficiency. As a result, only the hybridized potential $U^\mathrm{ZL}$ dominates, and the force is well approximated by:
\begin{equation}
	\mathbf{F}_{\mathrm{rad}} = - V_{\mathrm{p}}\nabla U \approx - V_{\mathrm{p}}\nabla U^\mathrm{ZL} = - V_{\text{p}} \nabla \left\langle U_{\text{inst}}^{\text{ZL}}\right\rangle,
\end{equation}
with $U_{\mathrm{inst}}^{\text{ZL}} = f_{1} p_{1}^{\mathrm{Z}} p_{1}^{\mathrm{L}} / (\rho_{0}c_{0}^{2}) - \frac{3}{2} f_{2}\rho_{0} \mathbf{v}_{1}^\mathrm{Z} \cdot \mathbf{v}_{1}^\mathrm{L}$. We note that $U_{\mathrm{inst}}$ is twice larger than in Eq.~\eqref{eq:u_inst} due to the binomial coefficients in the expansion of $p_{1}^{2}$ and $\mathbf{v}_{1} \cdot \mathbf{v}_{1}$.

\begin{figure}[!htbp]
	\includegraphics[width=\columnwidth]
	{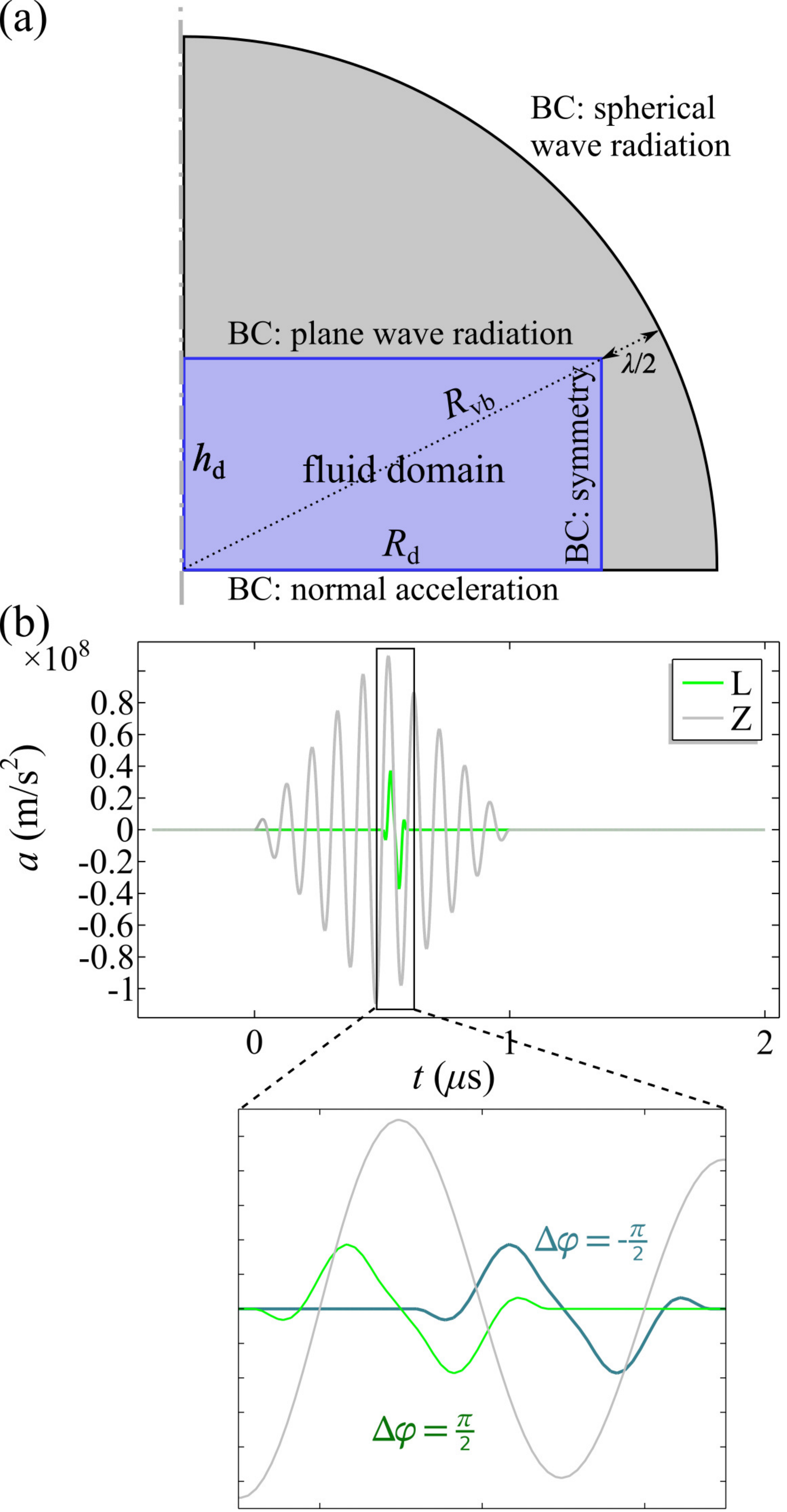}
	\caption{\label{fig:model} Model setup. (a) Sketch of the two-dimensional axial symmetry computational domain. The purple domain is used for the simulation of the Z-field and the calculation of the radiation force. The gray domain is used for the calculation of the L-field. (b) Input signals applied to the bottom boundary. Input signals are depicted by velocities varying with time. Grey and green curves represent Z-signal (electroacoustic BAW) and L-signal (photoacoustic pulse) respectively. The zoom-in shows the signal duration overlap under typical relative phases ($\Delta\varphi=\pm\pi/2$) within the time window of a period (from $0.47\weis$ to $0.62\weis$ ). }
\end{figure}

While we are mainly focused on the acoustic radiation pressure, we note that acoustic streaming plays a negligible role in this study due to the particle radius of $25\weim$ used in the experiments \cite{barnkob2012acoustic}. For fluids in the channel, thermophoresis and thermal agitation can also be factors driving the particles \cite{qiu2021fast}. To rule out this possibility, we have turned off the bulk acoustic wave (BAW) from Z-source and observed no motion of particles despite the laser heating \cite{wang2022laser}. Therefore, thermal effects are also neglected in the model. The other parameter values are given in Table~\ref{tab:parameter}. Most of these parameters are selected based on the typical experimental values while they are sometimes simplified to facilitate dimensionless analysis (for instance $f^{\text{Z}}$ is rounded to 10 MHz instead of the exact resonance frequency of the crystal used in experiments (7.6 MHz)). Further, an important goal of the paper is to estimate what would be the upper limit of the radiation force of the LGAT if suitable laser and electronic excitation were provided. For the electroacoustic field, a simple estimate is obtained by using the definition of the piezoelectric coupling coefficient $K^{2}$:
\begin{equation}
	\frac{1}{2}C(Ed)^{2}K^{2} \simeq \frac{1}{2}\rho_{\text{z}}Ad{v_{0}^{\text{Z}}}^{2},
\end{equation}
where $d$ is the thickness and $A$ is the area of the dielectric crystal, together with the capacitance $C=(\varepsilon_{\text{r}} \varepsilon_{0} A)/d$ and the field strength $E$. On this basis, an estimation of the velocity is derived as $v_{0}^{\text{Z}}=KE\sqrt{\varepsilon_{\text{r}} \varepsilon_{0}/\rho_{\text{z}}}$. This velocity is independent of the excitation frequency and is limited by the coercivity field strength of the crystal, which is of the order of \SI{21}{\mega\volt/m} for LiNbO\textsubscript{3} \cite{myers1995quasi}. The Y-36\textdegree\,cut has a $K^2=0.1$. For the sake of simplicity, the material anisotropy is overlooked and we consider a relative permittivity of the order of $\varepsilon_{\text{r}} \approx 40$, which yields a maximum velocity of \SI{1.83}{\m/s}.

To satisfy $U^{\text{LL}} \ll U^{\text{ZL}}$ as required by the hybridized-fields theory, the maximum excitation velocity of the photoacoustic field is set to \SI{60}{\cm/s}. We note that the material damage threshold could in principle allow higher velocities. More details about the criterion are given in Appendix \ref{app:ull&uzl}.

The laser spot being axisymmetric and the Z-wave being spatially-invariant, we use a two-dimensional axisymmetric model (cylindrical coordinates system) as shown in Fig.~\ref{fig:model}(a). We restrict our analysis to a rectangular cross-section (purple part in Fig.~\ref{fig:model}(a)) of width $R_{\text{d}}=n_{\lambda} \lambda^{\text{Z}}/2=375\weim$  (with $n_{\lambda}=5$) and height $h_{\text{d}}=200\weim$ in the vertical $r\text{-}z$ plane for a \SI{10}{\MHz} wave. The width is chosen large enough to ensure that the tweezers yield a negligible force at that distance.

In experiments, the particles are contained in a polydimethylsiloxane (PDMS) microchannel filled with water. PDMS and water have a similar acoustic impedance (plane wave reflection coefficient $R=18\%$). In order to reduce simulation time, the simulation domain size is minimized by neglecting acoustic reflections on the PDMS. Instead, plane wave radiation boundary conditions are used for the Z-wave, and spherical wave radiation ones are used for the L-wave. A virtual boundary of radius $R_{\text{vb}} = \sqrt{R_{\text{d}}^2+h_{\text{d}}^2}+\lambda^{\text{Z}}/2$ shown in Fig.~\ref{fig:model}(a) is constructed to facilitate the setting of spherical wave radiation boundary condition. More refined boundary conditions accounting for reflections in the PDMS-water surface and shear waves in the PDMS are discussed by Ni \textit{et al.} \cite{ni2019modelling}.

The transducer is not explicitly modeled, instead, the piezoelectric Z-excitation is implemented as an $r$-invariant acceleration:
\begin{equation}
	a^{\text{Z}} = a_{0}^{\text{Z}}{\sin\left( {\omega^{\text{Z}}t - \varphi^{\text{Z}}} \right)}\Lambda\left( \frac{\omega^{\text{Z}}t - \varphi^{\text{Z}}}{2\pi} \right).
\end{equation}
The triangular window $\Lambda(\cdot)$ with the trigger range of $[0,n^{\text{Z}}]$ simulates the resonance onset of the piezoelectric with a train of $n^{\text{Z}} = 5$ cycles, (see Fig.~\ref{fig:model}(b)). We note that longer excitation could be possible but would be more sensitive to reflections at the lateral edge of the wafer that would induce standing waves (and violate the in-plane invariance of the Z-wave). The relationship $\omega^{\text{L}} = \omega^{\text{Z}}$ exists unless specified otherwise.

Likewise, the photoacoustic excitation is described by an acceleration with a Gaussian intensity profile, since the photoacoustic vibration is well approximated by a Gaussian profile according to our experimental measurements \cite{wang2022optimization}. We set the radius of the beam waist $R_{\text{w}}$ close to the experimental value ($20\weim$), producing a normal acceleration given in Eq.~\eqref{eq:a_l}, where the amplitude reads $a_{0}^{\text{L}} = v_{0}^{\text{L}} / \omega^{\text{L}}$ . A rectangular window function modulates the sinusoidal signal to simulate the experimentally-detected pulsed L-signal (see Fig.~\ref{fig:model}(b)):
\begin{align}
	a^{\text{L}}  ={}& \frac{\partial v_{0}^{\text{L}}}{\partial t} \notag \\
	={}& a_{0}^{\text{L}}e^{- {({r/R_{\text{w}}})}^{2}}\frac{\partial}{\partial t}\left\lbrack {\sin\left( {\omega^{\text{L}}t - \varphi^{\text{L}}} \right)}\Pi\left( \frac{ \omega^{\text{L}}t - \varphi^{\text{L}} }{2\pi} \right) \right\rbrack. \label{eq:a_l}
\end{align}

Here the rectangular window $\Pi(\cdot)$ is set to 1 over the interval $[0,1/2]$ to simulate a pulse of duration $\tau^{\text{L}} = \pi/\omega^{\text{L}}$ (here $\tau^{\text{L}} = \SI{50}{\ns}$).

A smoothing of 0.4 is introduced to introduce frequencies above $f_{\mathrm{max}}=f^{\text{Z}}$ that would trigger numerical instabilities (see Appendix~\ref{app:smooth}). The smoothing process introduces small overshoots at the beginning and the end of the laser waveform function with an amplitude negligible compared to the main pulse peak ($\approx20\%$). Both excitations have an adjustable phase $\varphi$ that can be set in experiments using an adjustable delay to a common reference trigger. By default, $\varphi^{\text{Z}}=0$ and $\varphi^{\text{L}}=(2n^{\text{Z}}+1/2)\pi$ such that the two excitation waveforms are in-phase. When the phase between $Z$ and $L$ is adjusted, $\varphi^{\text{Z}}$ is the adjustment variable ($\varphi^{\text{L}}$ remains unchanged) and an empirical reference $\varphi_0(\mathbf{r})$ is introduced such that $\Delta\varphi = \varphi^{\text{Z}}- \varphi_0$ vanishes when the force is 0 at the measurement point $\mathbf{r}$. This will be discussed in more detail in section \ref{sec: effect of phase}.

\section{\label{sec:result}Results}
In this section, we illustrate the working of the laser-guided acoustic tweezers and discuss potential directions for parameter optimization. 
\subsection{Acoustic field}
As shown in Fig.~\ref{fig:model}(b), the simulation starts at $t = -0.2t_{\text{sim}}=-0.40\weis$. At $t=0$, the piezoelectric begins to emit a BAW (Z) wave. The photoacoustic pulse (L-signal) hits the substrate at $t=0.52\weis$. Therefore, the observation of time evolution begins at $t=0.52\weis$.

\begin{figure}[htbp]
	\includegraphics[width=\columnwidth]{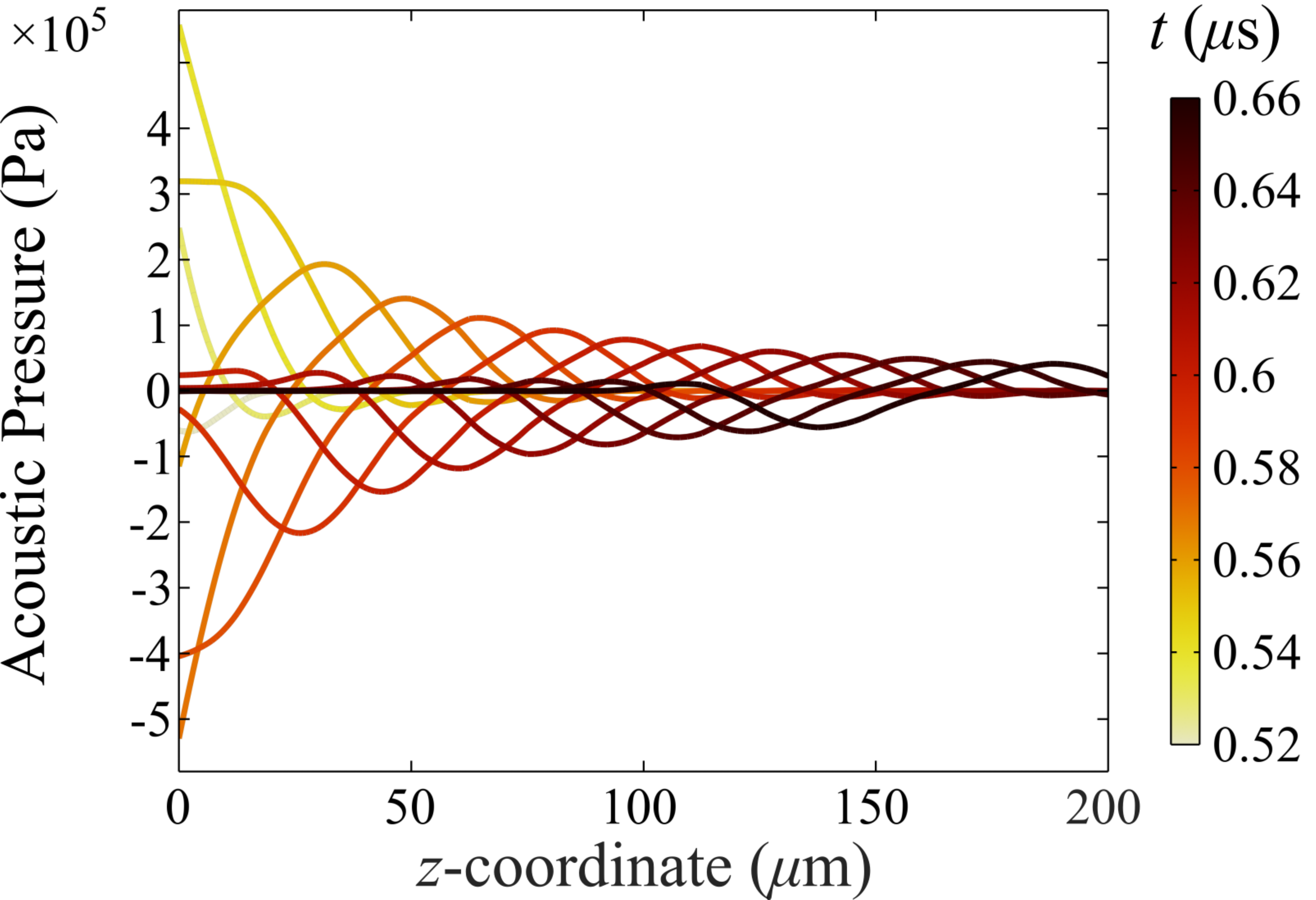}
	\caption{\label{fig:ap-z} Acoustic pressure of the L-field on the central axis of the simulation domain. As shown in Fig.~\ref{fig:model}, the laser beam hits the substrate at $t=0.52\weis$. }
\end{figure}
The Z-field is a plane traveling wave and remains similar to Fig.~\ref{fig:model}(b) as it travels in the $z$-direction. The L-wave generated by photoacoustic conversion is shown in Fig.~\ref{fig:ap-z}. It is essentially a spherical wave that behaves similarly in the $z$- and $r$-directions ($r$-profile available in Appendix ~\ref{app:ap-r}). The L-acoustic pressure reaches its maximum amplitude immediately after generation at $t=0.54\weis$  and decays by $80\%$ by the time it exists in the simulation domain ($t=0.66\weis$). 

\subsection{Acoustic radiation pressure}
In manipulation experiments, the particle is confined in the manipulation chamber. Therefore, radial forces tend to be more important than axial ones. Hence, unless specified otherwise, the ARF will refer to $F_{r}^{\text{ZL}} = - V_{\text{p}} \frac{\partial U^{\text{ZL}}}{\partial r}$. 

\subsubsection{Build-up of the Gor'kov potential}
\begin{figure*}[htbp]
	\includegraphics[width=\textwidth]{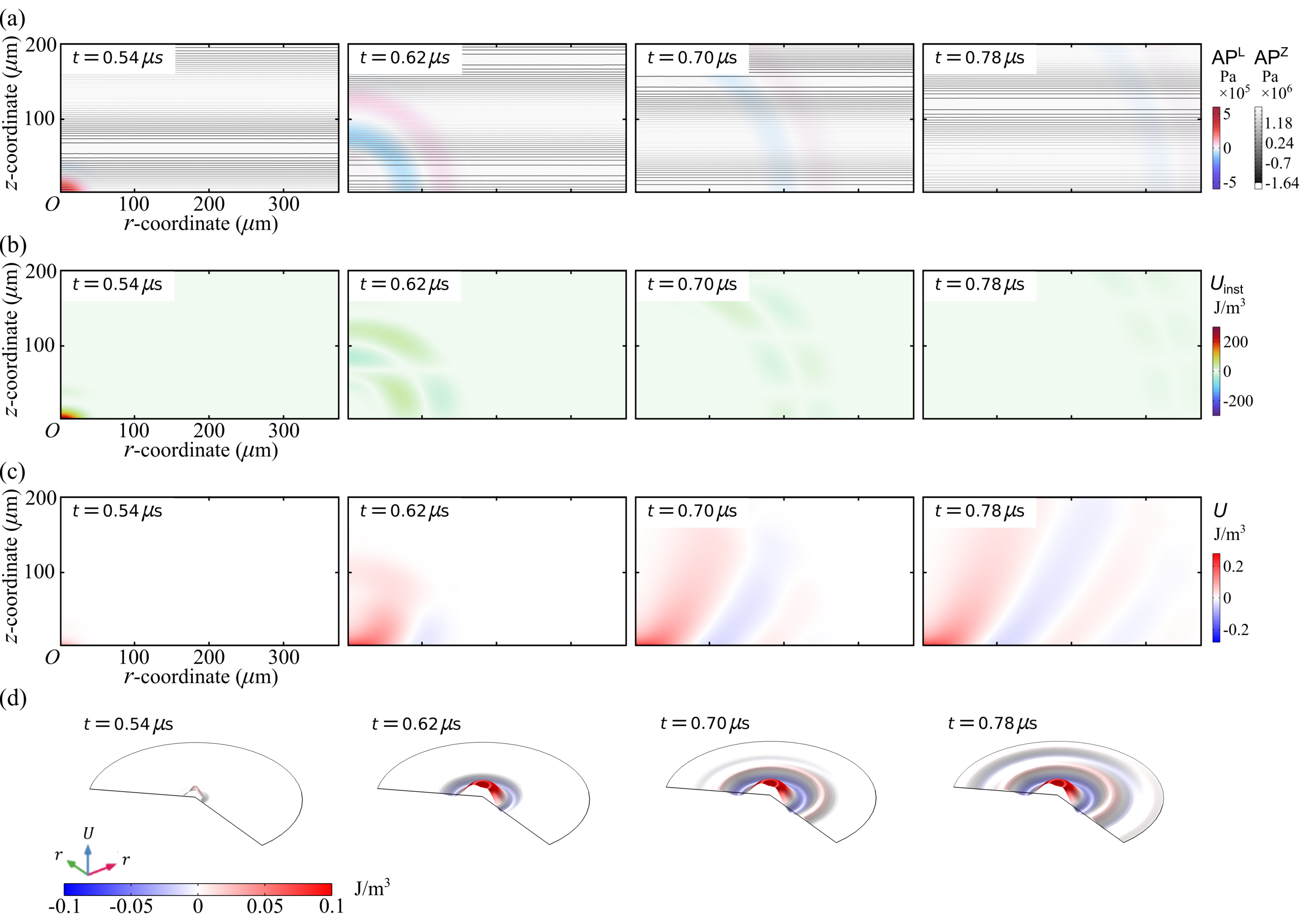}
	\caption{\label{fig:ap&u} Time-dependent field evolution with $\Delta\varphi=0$. Pictures are taken at $t=0.54$, 0.62, 0.70, and $0.78\weis$. (a) Super-imposed acoustic pressure field. The blue-red color represents the L-field pressure and the contour represents the Z-field pressure. (b) Instantaneous Gor'kov potential $U_{\text{inst}}$. (c) Time-averaged Gor'kov potential $U$. (d) Cross-section of the magnitude of $U$ at $z=R_{\text{p}}$.}
\end{figure*}

Fig.~\ref{fig:ap&u} shows the build-up of the acoustic radiation pressure. The pressure of the Z and L-waves are shown in Fig.~\ref{fig:ap&u}(a): the Z-wave is a traveling plane wave, and the L-wave is similar to a spherical wave. 

Despite being a virtual quantity, the instantaneous potential $U_{\text{inst}}$ to visualize the hybridization of the L and Z fields during the acoustic radiation pressure build-up. For example, the region where $U_{\text{inst}}<0$ in Fig.~\ref{fig:ap&u}(b) corresponds to the mixing of $p_{1}^{\text{L}}$ and $p_{1}^{\text{Z}}$ of opposite signs. This instantaneous Gor'kov potential grows over time, as shown in Fig.~\ref{fig:ap&u}(c). As the spherical wavefront (L) propagates, the inhomogeneous Gor'kov potential region spreads to the entire simulation domain. A cross-section of the hybridized Gor'kov potential $U^{\text{ZL}}$ experienced by a particle resting on the bottom wall ($z=R_{\text{p}}$) is shown in Fig.~\ref{fig:ap&u}(d). The potential is mainly concentrated at the location of the laser spot, which ensures good selectivity for acoustic trapping and manipulation. We note that our model neglects the effect of walls, which needs to be accounted for using a more advanced theory (see Ref.~\cite{baasch2020acoustic}  for the monofrequency case). 

\begin{figure}[htbp]
	\includegraphics[width=\columnwidth]{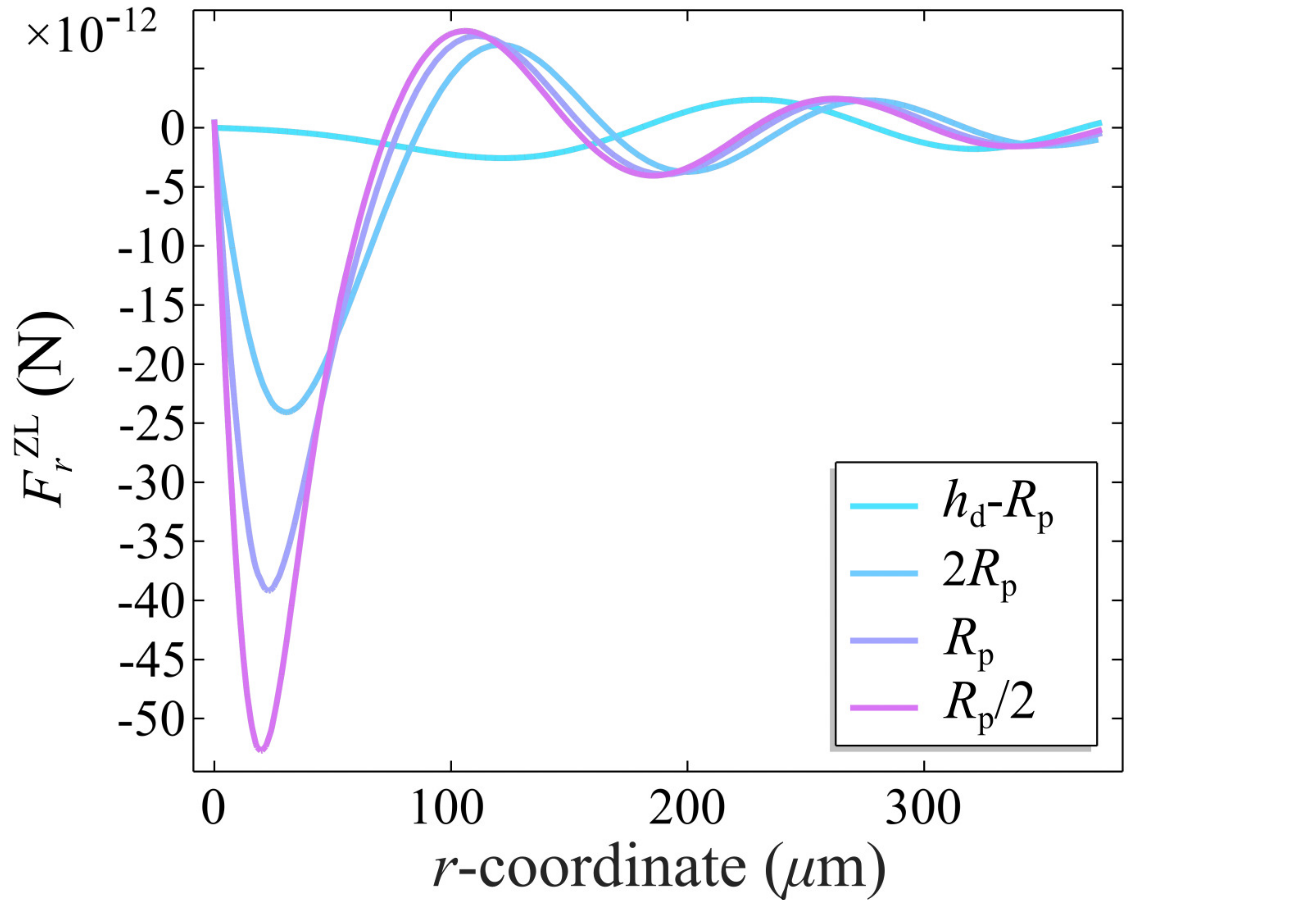}
	\caption{\label{fig:arf-z} Variation of the hybridized acoustic radiation force $F_{r}^{\text{ZL}}$ at various heights in the manipulation chamber when $\Delta\varphi=\pi/2$. }
\end{figure}

\subsubsection{Effect of the particle location}
The effect of the particle location is shown in Fig.~\ref{fig:arf-z}, with $\Delta\varphi=\pi/2$ selected to maximize the radiation force (this will be interpreted in section \ref{sec: effect of phase}). The closer the observation position $z$ is taken to the lower edge, the larger the ARF peak generated. The concentration of the pulse energy is also reflected in the ARF curves shown in Fig.~\ref{fig:arf-z}(a) and (b), where the ARF peaks show a decaying trend when moving away from the excitation source from either the $r$-direction or the $z$-direction. According to Fig.~\ref{fig:arf-z}, not only the force becomes very weak when the particle is located near the top of the channel, but the selectivity (ratio of primary force maximum to the secondary one) also deteriorates. It is therefore important in LGAT design to ensure that the particles are relatively close to the laser spot, which may limit the biocompatibility of the manipulation, unless holographic techniques are used \cite{wei2019generating}. In the following, the force is maximized by assuming that the particle rests at the bottom of the channel ($z=R_p$)

\subsubsection{Effect of the phase difference between Z and L fields \label{sec: effect of phase}}
\begin{figure}[htbp]
	\includegraphics[width=\columnwidth]{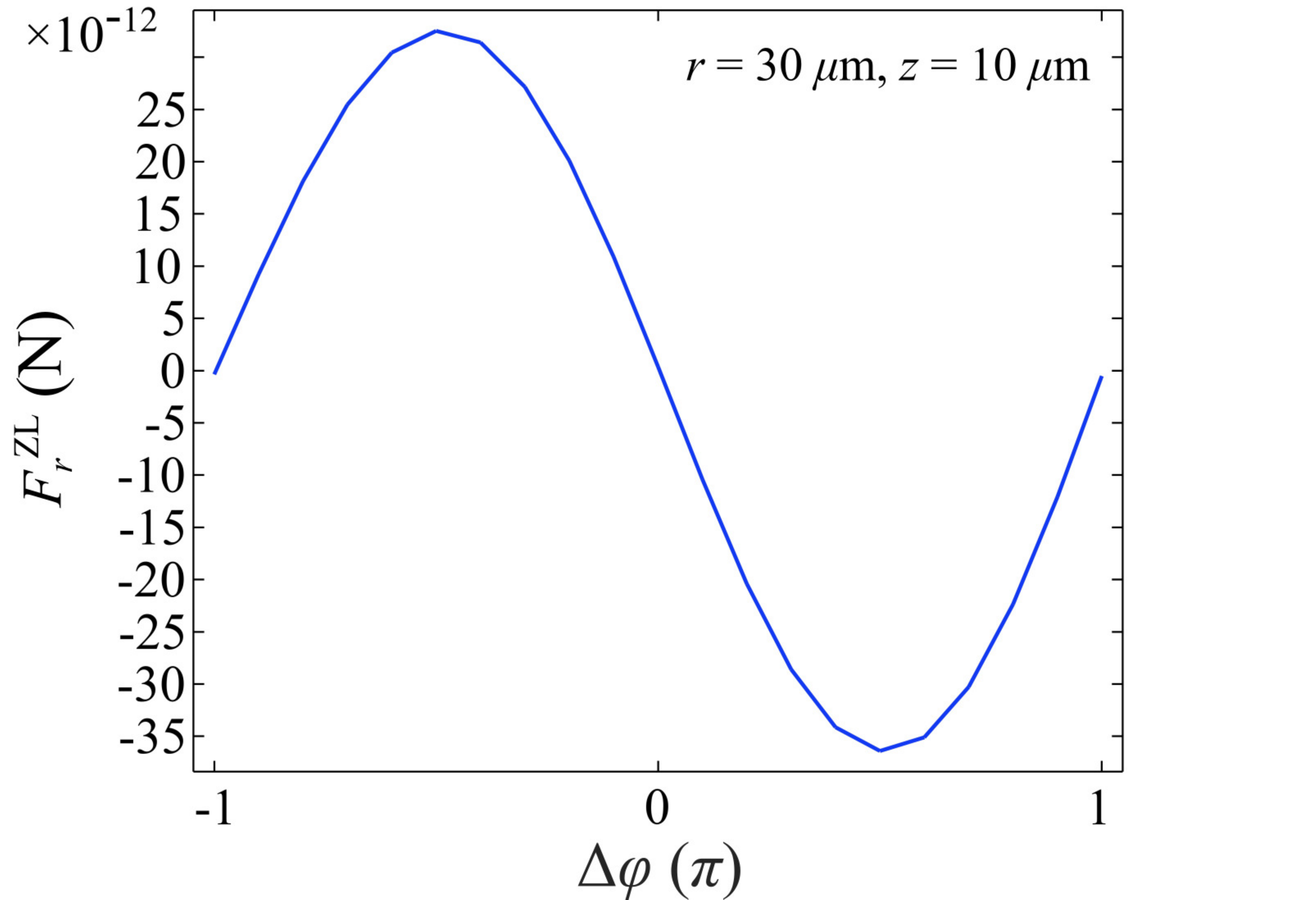}
	\caption{\label{fig:arf-phi} Hybridized acoustic radiation force $F_{r}^{\text{ZL}}$ at $z=R_{\text{p}}=10\weim$, $r=3R_{\text{p}}=30\weim$ depending the phase difference $\Delta\varphi$ between the Z and L fields. }
\end{figure}

A distinctive feature of LGAT is the possibility to commute between attractive and repelling forces by adjusting the electronic delay between the triggers of the laser source and the electric signal generator. In the simulation, this is implemented by changing the relative phase difference $\Delta\varphi$ between the Z and L fields, as shown for a particle located at $z=R_{\text{p}}=10\weim$, $r=3R_{\text{p}}=30\weim$ (Fig.~\ref{fig:arf-phi}). Following a convention proposed in our previous paper, we add an offset $\varphi_0(\mathbf{r})$ to $\Delta\varphi$ such that the ARF cancels for $\Delta\varphi=0$ \cite{wang2022laser}. A better alternative will be proposed later on. 

\begin{figure}[htbp]
	\includegraphics[width=\columnwidth]{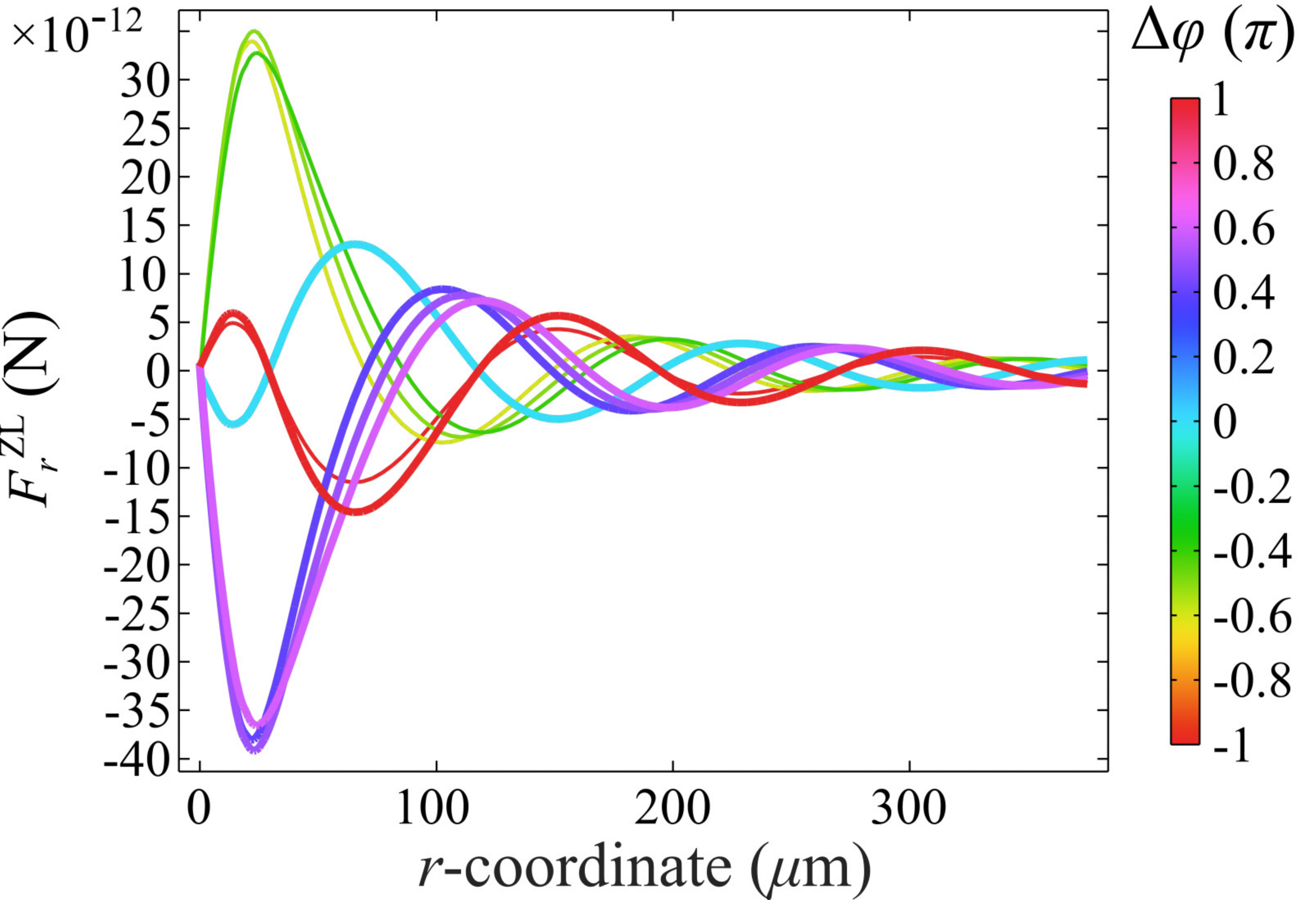}
	\caption{\label{fig:arf-r&phi} Variation of the hybridized acoustic radiation force $F_{r}^{\text{ZL}}$ at $z=R_{\text{p}}$ depending on the phase difference $\Delta\varphi$ between the Z and L fields. The line color indicates the phase while the line width grows with the phase in order to visually distinguish between $\Delta\varphi=-\pi$ and $\Delta\varphi=\pi$.}
\end{figure}

The variation of the ARF at fixed $z=R_{\text{p}}=10\weim$ but with a varying phase is shown in Fig.~\ref{fig:arf-r&phi}. Here, the phase offset is kept at $0$ such that $\Delta\varphi = \varphi^Z$ everywhere. Although at $r=3R_{\text{p}}=30\weim$ the force behaves consistently with Fig.~\ref{fig:arf-phi} (strongest pull at  $\Delta\varphi=\pi/2$, strongest push at  $\Delta\varphi=-\pi/2$ and vanishingly small at $\Delta\varphi=0$, it can be seen that points at half a wavelength away (such as $r=105\weim$) behave in the opposite fashion and points at a quarter wavelength distance (such as $r=62\weim$) behave in phase quadrature (strongest pull at $\Delta\varphi=\pi$, strongest push at  $\Delta\varphi=0$ and vanishingly small at $\Delta\varphi=\pm\pi/2$). This emphasizes that the exact delay between L and Z fields depends not only on the instruments trigger but also on the acoustic propagation of both fields. The latter is relative to the exact point in space where the ARF is measured. This issue is addressed in the next section. We also note that the force is not exactly periodic over time due to the slow variations of the wave envelope (the wave amplitude is not the same for $\Delta\varphi=-\pi$ and $\Delta\varphi=\pi$).

\subsubsection{A phase-independent factorization of the hybridized Gor'kov potential}
In the previous section, we have shown that the phase difference $\Delta\varphi$ is an important factor controlling the ARF of LGAT, but that the relationship between the force direction and $\Delta\varphi$ depends on the spatial location of the particle being manipulated. Hence, gaining a perfect knowledge of the ARF of an LGAT would require repeating ARF calculations for all phases $\Delta\varphi \in [-\pi,\pi]$. Here, we show how to reduce this large number of calculations to only two. Ref.~\cite{wang2022laser} gives a way to obtain the hybridized potential $U^{\mathrm{ZL}}$ by setting up an analytical Z-field wave with constant amplitude. Then $U^{\text{ZL}}$ is expressed as a linear combination of two basis potentials weighted by trigonometric functions of $\Delta\varphi$:
\begin{equation}
	\hat{U}^{\text{ZL}} = \Phi_{\text{c}}{\cos{\Delta\varphi}} +\Phi_{\text{s}} {\sin{\Delta\varphi}}, \label{eq:u_dec}
\end{equation}
with the basis potentials 
$\Phi_{\text{c}}\left( \mathbf{r} \right) = \left\langle {p_{1}^{\text{Z}}|_{\Delta\varphi = 0}E^{\text{L}}} \right\rangle$ and 
$\Phi_{\text{s}}\left( \mathbf{r} \right) = \left\langle {p_{1}^{\text{Z}}|_{\Delta\varphi = \pi/2}E^{\text{L}}} \right\rangle$ independent of $\Delta\varphi$, where L-component 
$E^{\text{L}} = f_{1} p_{1}^{\text{L}}/(\rho_{0}c_{0}^{2}) - \frac{3}{2}f_{2} v_{1z}^{\text{L}}/c_{0}$. Here only the $z$-direction component of $\mathbf{v}_{1}^\text{L}$ is considered due to $z$-propagating-only $\mathbf{v}_{1}^\text{Z}$. $\hat{U}^{\text{ZL}}$ obtained from Eq.~\eqref{eq:u_dec} is strictly valid only for monofrequency Z-wave but is assumed here to be a good approximation of the actual potential $U^{\text{ZL}}$. In the simulations,  $\Phi_{\text{c}}$ ($\Phi_{\text{s}}$) is obtained by setting $\Delta\varphi = 0$ ($\Delta\varphi = \pi/2$). The approximate potential can then be computed from these two basis potentials according to Eq.~\eqref{eq:u_dec}. 

\begin{figure}[htbp]
	\includegraphics[width=\columnwidth]{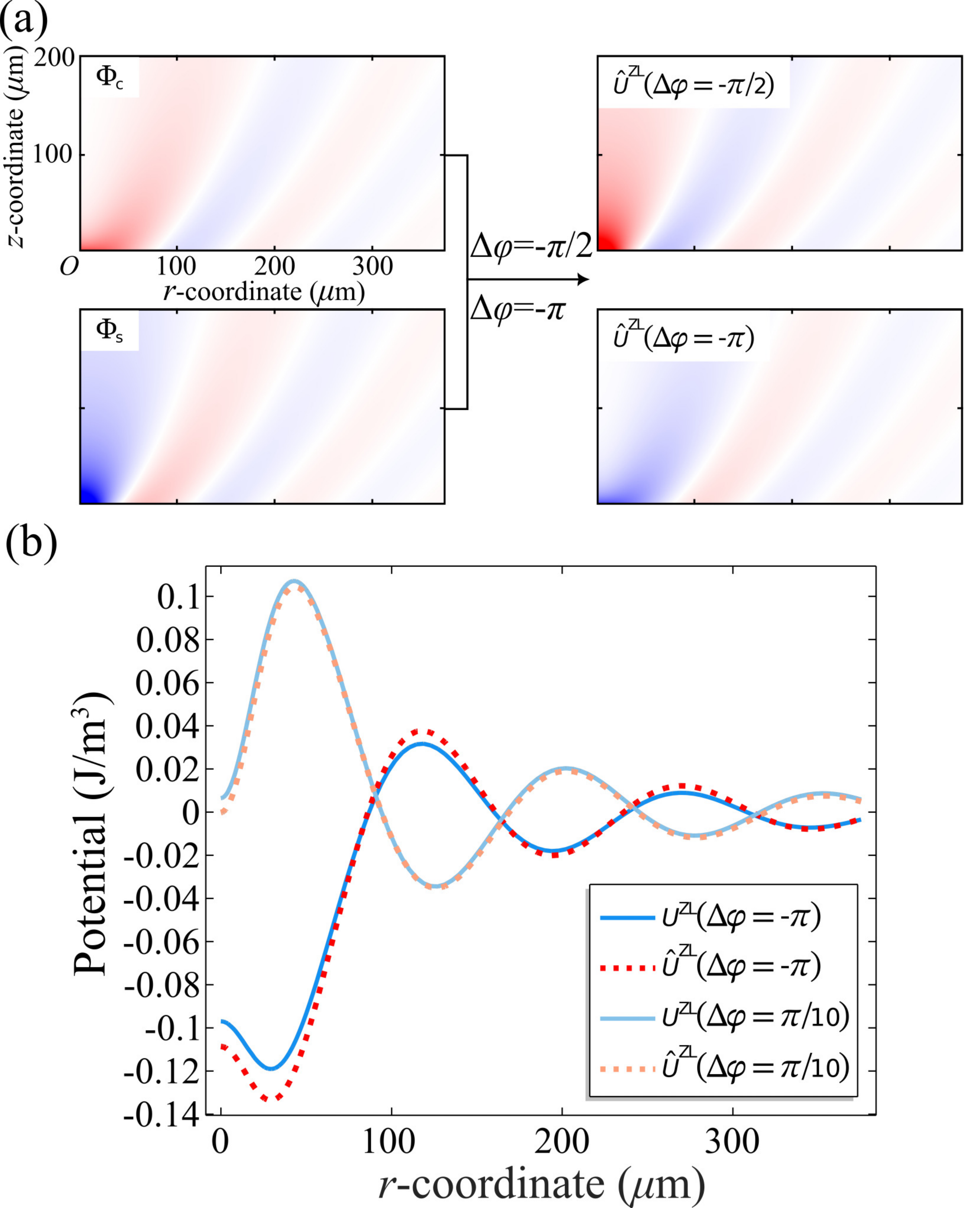}
	\caption{\label{fig:u_dec} Approximate Gor'kov potential $\hat{U}^{\text{ZL}}$ with $\Delta\varphi$ decoupled (a) Constant basis potentials $\Phi_{\text{c}}$ and $\Phi_{\text{s}}$. (b) Comparison between $\hat{U}^{\text{ZL}}$ and $U^{\text{ZL}}$ for $\Delta\varphi = -\pi$ and $\pi/10$.}
\end{figure}

Fig.~\ref{fig:u_dec}(a) shows the basis potentials at $z=R_{\text{p}}$. Their values can be recorded in the entire simulation domain. The quality of the approximation is evaluated in Fig.~\ref{fig:u_dec}(b): the approximated potential matches $U^{\text{ZL}}$ (from Eq.~\eqref{eq:u_avg}) well, although the relative error for $\Delta\varphi = -\pi$ increases by $\sim 8\%$ compared to the one of $\Delta\varphi = \pi/10$. This is likely because the Z-wave has a triangular envelope of varying amplitude which is not considered in the theory of Ref. \cite{wang2022laser}.
 
\subsection{Optimization of operating parameters}
The main weakness of current LGAT implementations is that they yield a very small force that results in displacement speeds of a few micrometers per second. In the following, we optimize two parameters readily adjustable in experiments: (1) the ratio of the L-wave frequency (reflecting the laser pulse duration) with respect to the Z-wave frequency, i.e., $\xi_{f} = f^{\text{L}}/f^{\text{Z}}$ (which can be adjusted on most pulsed laser sources); (2) the size of the pulse source, expressed as the radius of the laser spot $R_{\text{b}}$ (which can be adjusted by changing the microscope objective magnification). We assume that $\xi_{f}$ is adjusted by changing $f^{\text{L}}$ with $f^{\text{Z}}$ fixed at \SI{10}{\MHz}. In this section, the conditions $\Delta\varphi = \pi/2$ and $z=R_{\text{p}}$ are retained, and the radiation force magnitude ${F_r}^{\text{max}} $ is discussed based on the maximum absolute values of the ARF in the $r$-direction. 

To reflect experimental constraints, the comparisons are done at constant laser peak energy (that would be equivalent to adjusting the laser energy to avoid material damage). Photoacoustic generation theory shows that the photoacoustic wave amplitude is proportional to the laser power. Hence, we adjust the photoacoustic wave amplitude as $v_{0}^{\text{L}}\xi_{f}$ (see Appendix \ref{app:vscale}). In addition, the laser phase $\varphi^{\text{L}}$ is adjusted to keep the peak aligned with the electroacoustic field peak.

\begin{figure}[htbp]
	\includegraphics[width=\columnwidth]{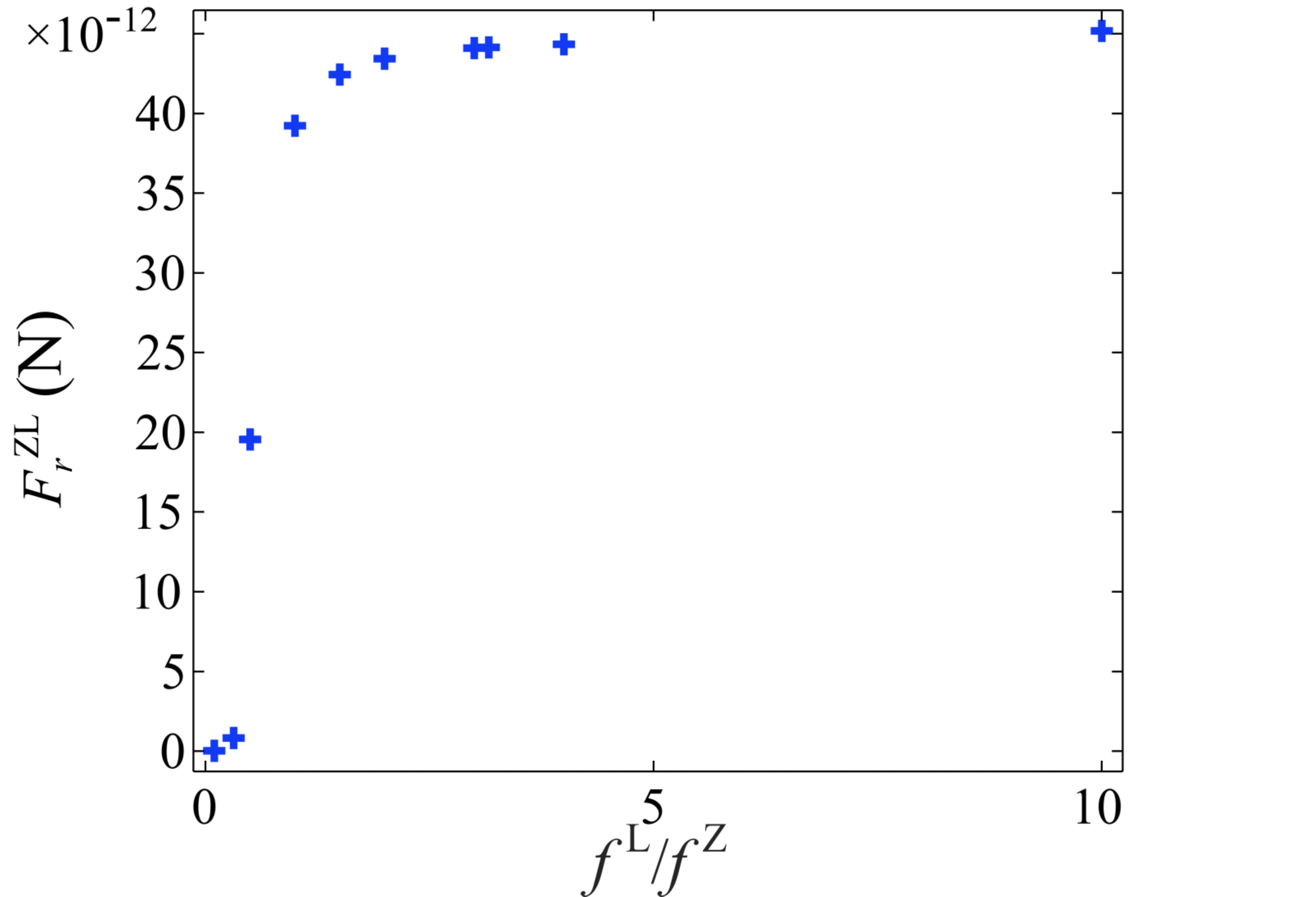}
	\caption{\label{fig:arf-f} Hybridized acoustic radiation force  $F_{r}^{\text{ZL}}$ for various pulse duration $(f^{\text{L}}/f^{\text{Z}})$ while keeping the laser pulse energy constant.}
\end{figure}

Fig.~\ref{fig:arf-f} illustrates the effect on ARF of modulating $\xi_{f}$. The ARF increases fast as the pulse shortens until it reaches a saturation value of ${F_r}^{\text{max}}\sim\SI{44}{\pico\newton}$ when the pulse frequency satisfies ($f^{\text{L}} \sim f^{\text{Z}}$). Afterward, the increase flattens. While the marginal improvement past $f^{\text{L}} \sim f^{\text{Z}}$ might seem attractive for improved versions of the LGAT, we note that the short wavelength at high frequencies violates Gor'kov's assumption of an object size much smaller than the wavelength, and the onset of resonance can yield to an uncontrollable force direction.

Next, we consider varying the laser spot size, for instance by using objectives of varying magnification. Here, $R_{\text{b}}$ is introduced to distinguish the adjusted spot radius from the original radius $R_{\text{w}}$ used in the other parts of the paper. The ratio $\xi_{R}=R_{\text{b}}/R_{\text{w}}$  is defined for convenience. To maintain constant pulse energy, the scaling $v_{0}^{\text{L}}/\xi_{R}^{2}$ neglecting the loss of transmission of high-magnification objectives is used (detailed derivation in Appendix \ref{app:vscale}).

Fig.~\ref{fig:arf-rb}(a) and (b) show the evolution of the ARF depending on the pulse source radius when $\Delta\varphi = 0$ and  Fig.~\ref{fig:arf-rb}(c) and (d) shows the evolution of the ARF depending on the pulse source radius when $\Delta\varphi = \pi/2$. It is found that $\Delta\varphi = \pi/2$ yields the largest force regardless of the frequency and beam radius. Optimizing the latter two parameters ($f^{\text{L}}=\SI{14}{\MHz}$ (pulse duration of \SI{31}{\ns}), $R_{\text{b}}= 10\weim$), we expect a force of $\sim \SI{80}{\pico\newton}$, well within the range of other acoustic tweezers, but that would have the extra advantage of being controlled by a light pattern.

\begin{figure*}[htbp]
	\includegraphics[width=\textwidth]{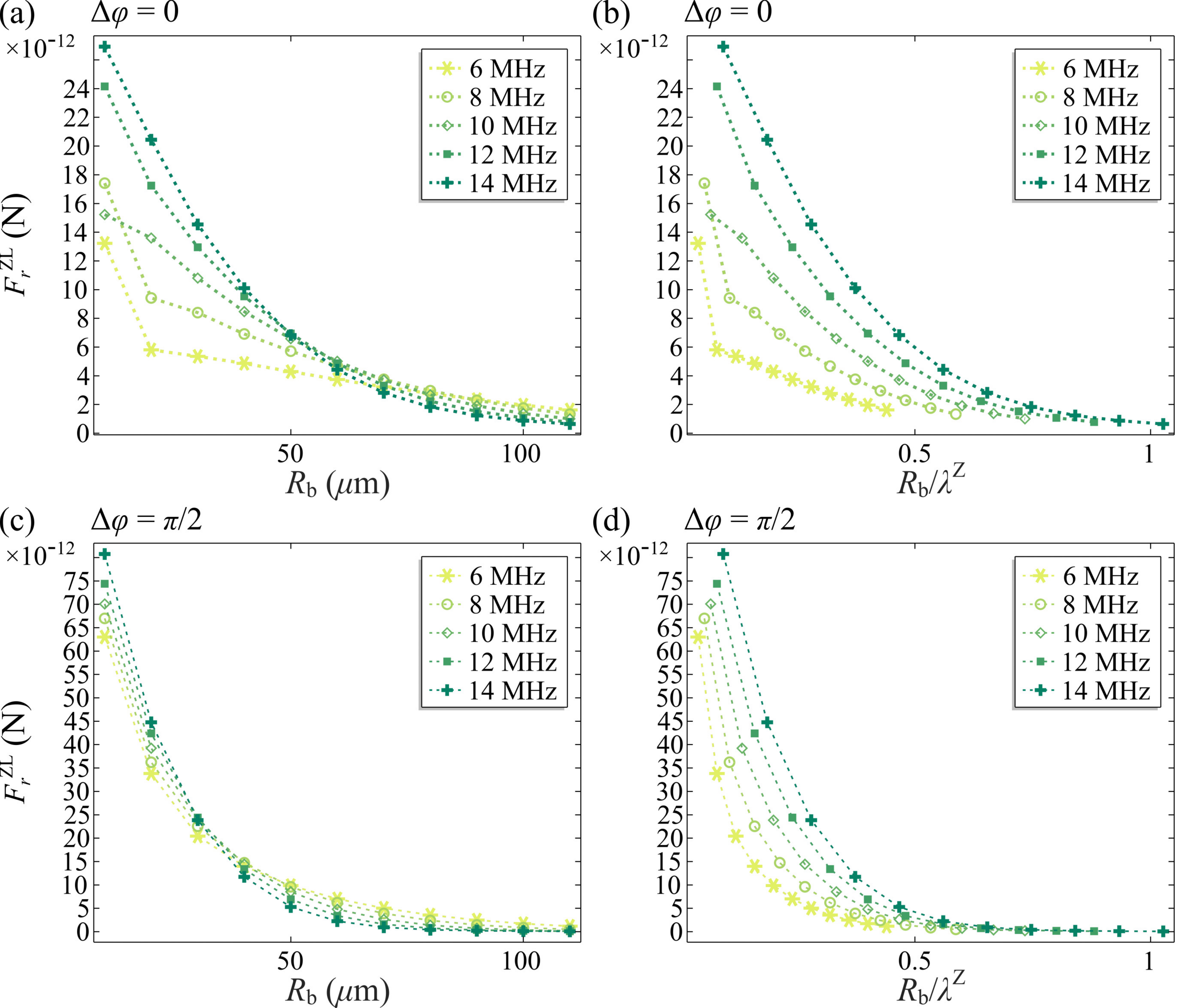}
	\caption{\label{fig:arf-rb} Hybridized acoustic radiation force  $F_{r}^{\text{ZL}}$ for various pulse durations (expressed as frequencies of 6, 8, 10, 12, and \SI{14}{\MHz}) and beam radii, while keeping the beam energy constant. (a) and (b) presents the force versus the  beam radius $R_{\text{b}}$  and the dimensionless radius $R_{\text{b}}/\lambda^{\text{Z}}$ for $\Delta\varphi = 0$ while (c) and (d) presents those for $\Delta\varphi = \pi/2$.}
\end{figure*}

\section{\label{sec:perspective}Perspective}
In the model, we have only discussed the case of small particles in an ideal fluid domain, and the properties of the wave sources are relatively simplified. Therefore, the model deserves further extensions, such as the introduction of thermoviscous effects and interaction with walls. Furthermore, we have assumed in the model that the radius of the particle was small enough to be consistent with the Gor'kov approximation \cite{hasegawa1977comparison}. However, the LGAT experiments \cite{wang2022laser} of manipulation of $\sim 25\weim$ radius particles have unveiled a much larger force, presumably due to the onset of resonance. Therefore, a theory for transient ARF valid for arbitrary particle radii would be highly desirable. 

\section{\label{sec:conclusion}Conclusion}
Despite the availability of a theoretical expression for the ARF of short pulses, no finite element models had been implemented so far. In this paper, we have shown that such a simulation could be carried out by (i) using time-explicit acoustic pressure simulations to compute the acoustic field, (ii) defining a virtual instantaneous Gor'kov potential, and (iii) integrating it over time to obtain the time-averaged Gor'kov potential. We have then used our code to simulate laser-guided acoustic tweezers. Our simulations show that the behavior of the LGAT can be captured by two conjugated Gor'kov potentials $\Phi_{\text{c}}$ and $\Phi_{\mathrm{s}}$ weighted by trigonometric functions. Optimization of operating parameters shows that the force increases when reducing the laser spot size, while the gain from shortening the pulse duration is limited ($\sim 10\%$). Finally, based on the dielectric breakdown limit of the piezoelectric, we estimate the maximum LGAT force on $10\weim$ polystyrene particles in water to $\sim\SI{40}{\pico\newton}$, which is comparable to other acoustic tweezers. Beyond LGAT, the availability of a numerical model may accelerate the development of time-dynamic acoustic tweezers with superior dexterity compared to their monochromatic counterpart.

\begin{acknowledgments}
This work was supported by the National Natural Science Foundation of China with Grants Nos. 12004078, 61874033, and 62274039; the State Key Lab of ASIC and System, Fudan University with Grants Nos. 2021MS001, 2021MS002, and 2020KF006; and the Science and Technology Commission of Shanghai Municipality with Grants Nos. 22QA1400900 and 22WZ2502200.
\end{acknowledgments}
\appendix
\section{\label{app:ull&uzl} Magnitude comparison between \texorpdfstring{\textit{U}\textsuperscript{LL}}{ULL} and \texorpdfstring{\textit{U}\textsuperscript{ZL}}{UZL}}
The simulation method follows the hypothesis of the hybridized-fields theory \cite{wang2022laser}, which expects a relatively low value of $U^{\text{LL}}$ compared to $U^{\text{ZL}}$. $U^{\text{LL}}$ is independent of $\Delta\varphi$. Here we evaluate the effect of $U^{\text{LL}}$ with the maximum $U^{\text{ZL}}$ induced by adjusting $\Delta\varphi$.

\begin{figure}[htbp]
	\includegraphics[width=\columnwidth]{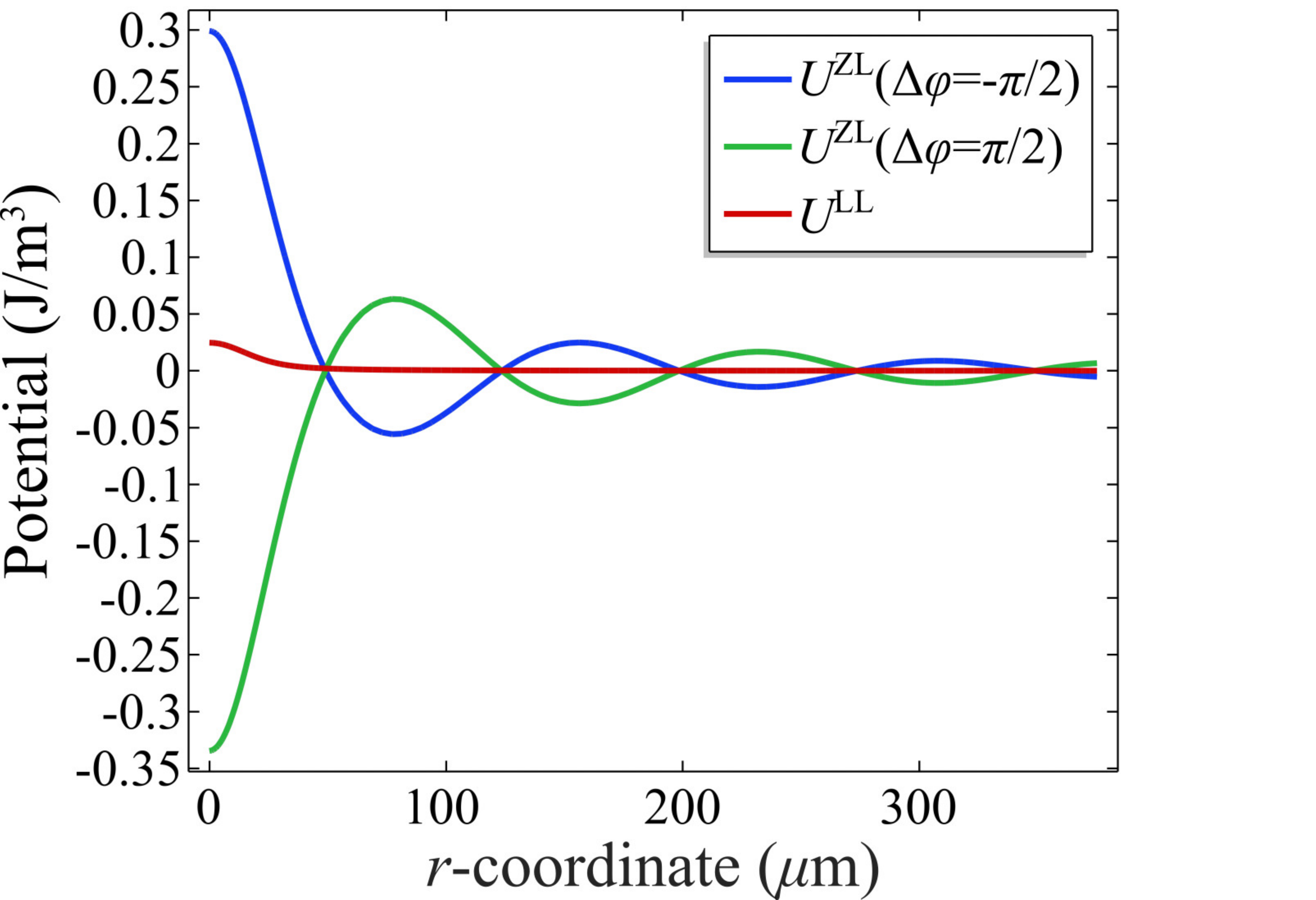}
	\caption{\label{fig:ull&uzl} Magnitude comparison between $U^{\text{LL}}$ and $U^{\text{ZL}}$ with $\Delta\varphi=\pm\pi/2$. }
\end{figure}

As shown in Fig.~\ref{fig:ull&uzl}, the maximum $U^{\text{LL}}$ reaches less than $10\%$ of the maximum $U^{\text{ZL}}$, which validates our assumption for the current simulation settings.

\section{\label{app:smooth} Effects of input signal smoothing}
\begin{figure}[htbp]
 	\includegraphics
 	[width=\columnwidth]
 	{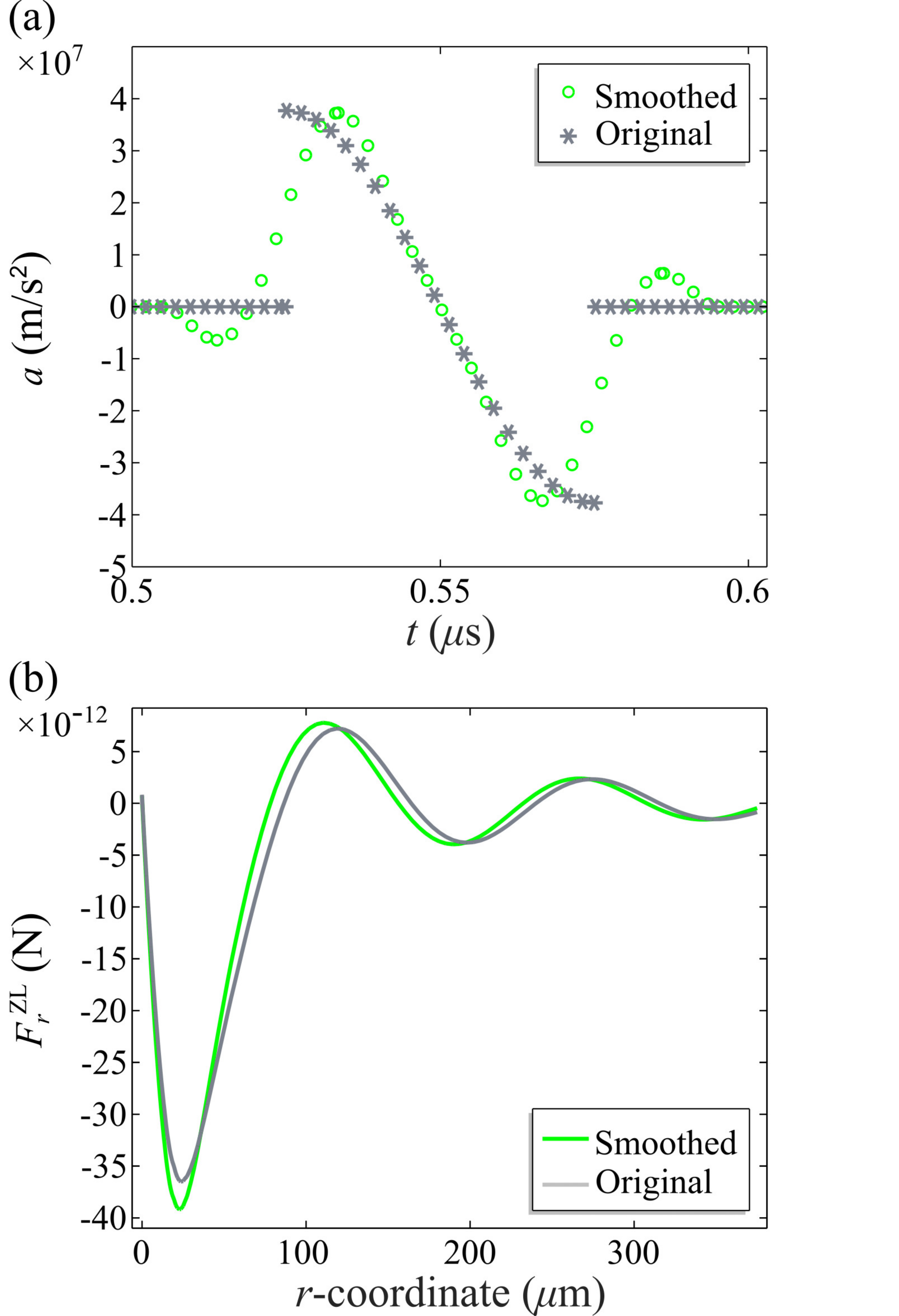}
 	\caption{\label{fig:smooth} Comparison between the sharp signal and the smoothed signal. (a) Signal input. (b)  $F_{r}^{\text{ZL}}$ on $z=R_{\text{p}}$ with $\Delta\varphi=\pi/2$. }
\end{figure}

Since the acceleration is derived from the derivation of the velocity, we will get an input $a^{\text{L}}$ with a sharp variation at the beginning and the end of the signal (as shown in Fig.~\ref{fig:smooth}(a)). Large changes tend to generate numerical errors. To minimize the issue, we introduce a transition region of width $0.4\omega^{\text{L}}/2\pi$ on each side of the rectangular window. In Fig.~\ref{fig:smooth}(b), we check that the deviation of the ARF resulting from the smoothing is moderate when compared to the non-smoothed function.

\section{\label{app:ap-r} Isotropy of the laser pulse}
\begin{figure}[!htbp]
	\includegraphics[width=\columnwidth]{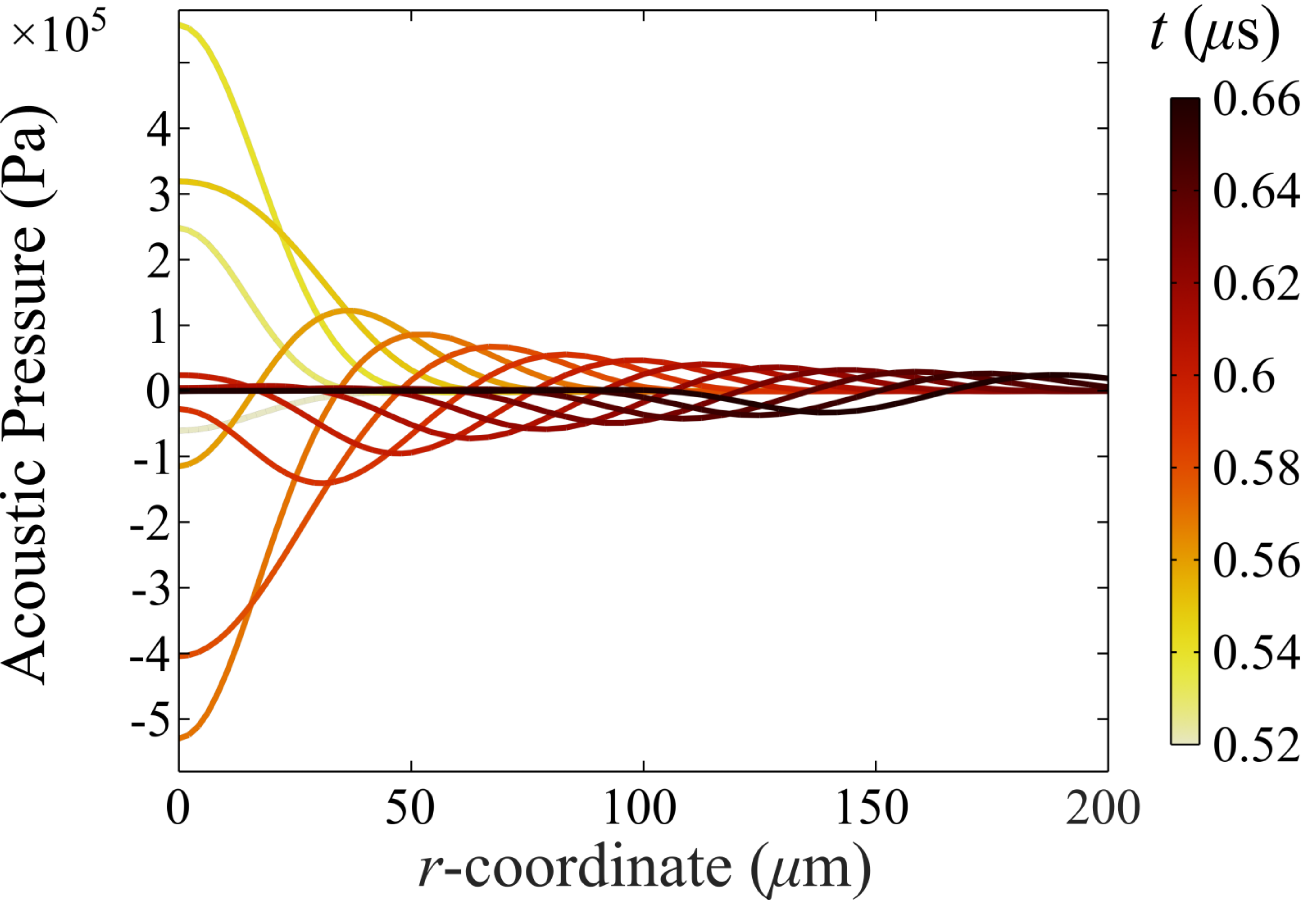}
	\caption{\label{fig:ap-r} Propagating properties of L-acoustic pressure varying with $r$ on the lower edge ($z=0$). The color bar from light yellow to dark red represents the change of time from $0.52\weis$ to $0.66\weis$.}
\end{figure}

The consistency of the variation trend and magnitude of the acoustic field in the $r$-direction (Fig.~\ref{fig:ap-r}) with that in the $z$-direction (Fig.~\ref{fig:ap-z}) reflects the spherical wave approximation condition. However, in the near-field zone, the acoustic pressure change in the $r$-direction is sharper. Such difference in propagation characteristics of different directions originates from the normal direction of the initial acceleration, while the rapid attenuation of acoustic fields in both directions reflects the strong concentration of acoustic energy obtained by photoacoustic generation.

\section{\label{app:vscale} Scaling velocity to maintain a constant pulse energy}
When changing operating parameters in the simulations, the laser pulse energy is assumed to remain constant. This is achieved by scaling the L-velocity according to the Vashy-Buckingham Pi-theorem. The velocity is assumed to read:
\begin{equation}
v^{\text{L}}(f^{\text{L}} t,r/R_{\text{b}}) = v^{\text{L}}_0 (\xi_R,\xi_f) g(f^{\text{L}} t, r/R_{\text{b}}), \label{eq: scaled v_L}
\end{equation}
where $g$ is an arbitrary integrable function that can be obtained in our simulations by integrating Eq.~\eqref{eq:a_l}. The goal of the following calculation is to determine $v^{\text{L}}_0(\xi_R,\xi_f)/v^{\text{L}}_0(1,1)$.

The pulse energy from the laser reads $E = \int_{A_{\mathrm{tot}}}\int_{t_{\mathrm{tot}}} \varepsilon \dif S \dif t $, where $\varepsilon$ is the pulse energy density. Substituting $\varepsilon=\beta^{-1}v^{\text{L}}$, with $\beta$ the photoacoustic conversion coefficient (assumed to be constant for the limited frequency and power range studied here) and using Eq.~\eqref{eq: scaled v_L}, we get:
\begin{equation}
E \approx \beta^{-1} v^{\text{L}}_0 (\xi_R,\xi_f) \int_\infty \int_\infty g(f^{\text{L}} t, r/R_{\text{b}}) \dif S \dif t. \label{eq: pulse energy}
\end{equation}
The right-hand side integral is obtained by assuming that the simulation domain is large enough to completely encompass the laser spot. Changing the integration variables to $f^{\text{L}} t$ and $r/R_{\text{b}}$, we get:
\begin{equation}
E \approx  \beta^{-1} v^{\text{L}}_0 (\xi_R,\xi_f) \frac{{R_{\text{b}}}^2}{f^{\text{L}}} I_g, \label{eq: integ v_L}
\end{equation}
where $I_g$ is the definite integral of the $g$ function.

A similar development at $(\xi_R,\xi_f)=(1,1)$ yields:
\begin{equation}
E \approx  \beta^{-1} v^{\text{L}}_0(1,1) \frac{{R_{\text{w}}}^2}{f^{\text{Z}}} I_g, \label{eq: integ v_L2}
\end{equation}
which yields $v^{\text{L}}_0 (\xi_R,\xi_f)/v^{\text{L}}_0 (1,1) = \xi_f/{\xi_R}^2$ that was used in the simulations.

\bibliography{LGAT-sim-ref-ab}

\end{document}